\documentclass[showpacs,preprintnumbers,amsmath,pre,amssymb]{revtex4}

\usepackage{graphicx}
\usepackage[latin1]{inputenc}
\usepackage{dcolumn}
\usepackage{bm}
\usepackage{epsfig}
\usepackage{amssymb, amsmath}

\begin{document}

\title{Horizontal Visibility graphs generated by type-II intermittency}

\author{\'Angel M. N\'u\~{n}ez}
\affiliation{Dept. Matem\'{a}tica Aplicada y Estad\'{i}stica. ETSI Aeron\'{a}uticos, Universidad Polit\'{e}cnica de Madrid, Spain.}
\author{Lucas Lacasa}
\affiliation{School of Mathematical Sciences, Queen Mary University of London, Mile End Road, London E1 4NS, UK}
\author{Jose Patricio G\'{o}mez}
\affiliation{Dept. Matem\'{a}tica Aplicada y Estad\'{i}stica. ETSI Aeron\'{a}uticos, Universidad Polit\'{e}cnica de Madrid, Spain.}

\date{\today}


\begin{abstract}
In this contribution we study the onset of chaos via type-II intermittency within the framework of Horizontal Visibility graph theory. 
We construct graphs associated to time series generated by an iterated map close to a Neimark-Sacker bifurcation and study, 
both numerically and analytically, their main topological properties. We find well defined equivalences between the main statistical 
properties of intermittent series (scaling of laminar trends and Lyapunov exponent) and those of the resulting graphs, and accordingly 
construct a graph theoretical description of type-II intermittency. We finally recast this theory into a graph-theoretical 
renormalization group framework, and show that the fixed point structure of RG flow diagram separates regular, critical and chaotic dynamics.
\end{abstract}

\maketitle

\section{Introduction}

Some relatively recent strategies in nonlinear analysis techniques are based 
on the mapping of time series into graphs according to different algorithms 
and criteria (see for instance \cite{lacasaetal2008,zhang06,kyriakopoulos07,xu08,donner10,campanharo11}) 
and on the subsequent study of the associated graphs. Topological features 
of the associated graphs are then related to dynamical aspects of the 
systems that generated the series and these methods are used for feature classification based analysis of complex signals.\\
Amongst them, the so called Horizontal Visibility Algorithm (HVA) \cite{luqueetal2009, lacasaetal2008} 
can be distinguished from others by being capable of distilling time series, 
condensing classes of them into a single graph whose structural properties represent 
the basic common dynamical properties of the series. The method 
uncovers structural features and forms sets of time series with the same feature by their 
representative HV graph ensemble, excluding from the ensemble those that lack that feature. 
The kernel dynamics in each case is well captured by the associated graphs, such that 
when the HV method is applied to a time series of unknown source, inspection of the resulting 
graph provides basic information about its underlying dynamics. Some relevant applications of 
this approach include the discrimination of reversible from irreversible dynamics \cite{nunhez12}, 
the characterization of chaotic and stochastic signals \cite{luqueetal2009, toral10} or, in general, 
applications to series classification problems where the HV is used as the feature extraction method 
(see \cite{review_graphs} for a recent review). However, from a theoretic point of view, the method 
is still in its infancy and graph-theoretical descriptions of nontrivial dynamics are in general 
open problems. This is specially important within this methodology, which has been proved to be 
simple enough to be addressed analytically instead of being yet another (black box) classification 
method, while at the same time being accurate and powerful for series classification.\\

\noindent In the context of low-dimensional chaos, two of the canonical routes to chaos 
(Feigenbaum scenario and quasiperiodic route) have been studied from this perspective 
and complete sets of graphs that encode the dynamics of their corresponding classes of 
iterated maps have been introduced and characterized \cite{plos, quasi}. The third 
canonical route to chaos is the so called Pomeau-Manneville or intermittency route 
\cite{schuster}. Under the generic term intermittency several dynamical behaviours 
with a common feature can be considered. The common feature is the alternation 
of (pseudoperiodic) laminar episodes with sporadic break-ups or bursts between 
them called intermissions. Intermittent behaviour can indeed be observed experimentally 
in many situations such as Belousov-Zhabotinski chemical reactions, Rayleigh-Benard 
instabilities, or turbulence \cite{maurer, pomeau, berge, schuster} and has been deeply 
studied in the context of nonlinear sciences. As a result of this, a characterization 
of the onset mechanisms and main statistical properties of intermittency have been 
described and typified: from the classification of types I, II and III intermittency 
by Pomeau and Manneville \cite{pom-mann} to other more recent types such as on-off 
intermittency \cite{platt} or ring intermittency \cite{hramov}. \\
The theoretical description of type-I intermittency from a Horizontal Visibility 
perspective has been advanced recently  \cite{interm_I}. In the present contribution 
we extend the HV description to type-II intermittency and present the structural, 
scaling and entropic properties of the graphs obtained when the HV formalism is 
applied to the type-II intermittency case, further advancing the HV theory. 
We recall here that type-II as described 
by Pomeau and Manneville in their seminar paper \cite{pom-mann} is not only of theoretical 
interest, but indeed constitutes a physical mechanism which has been experimentally identified, 
for example, in coupled nonlinear oscillators \cite{nonlin_osc} or hydrodinamic systems \cite{hyd_sys}.\\

\noindent In the following we first recall in section II the key aspects of type-II intermittency. 
In section III we outline the basic methodology defined as 
the Horizontal Visibility algorithm and apply it to the study of trajectories generated 
by iterated maps close to a Neimark-Sacker bifurcation, where type-II intermittency 
takes place. An heuristic derivation of an analytical expression for the degree distribution 
$P(k;\epsilon)$ of this kind of graphs is performed. This graph measure 
encodes the key scaling property of type-II intermittency: the mean length 
$\langle \ell \rangle $ of the laminar episodes with $\epsilon$ manifests in network realm 
as a comparable scaling with the same variable of the second moment $\langle k^{2}\rangle$ 
of the degree distribution $P(k,\epsilon)$. In turn, the scaling of Lyapunov exponent 
$\lambda(\epsilon)$ is recovered in network space, via Pesin-like identity, 
from Shannon block entropies $h_n$ over 
$P(k_1,k_2,...,k_n;\epsilon)$, whose block-$1$ entropy $h_1$ is only a first order approximation. 
In section IV we recast the family of HV graphs generated by intermittent 
series into a graph-theoretical Renormalization Group (RG) framework and determine the RG
flows close to and at the bifurcation point. As in other transition-to-chaos 
scenarios \cite{plos,quasi,interm_I}, we find two trivial fixed points of the RG flow 
(akin to the high and low temperature fixed points 
in thermal phase transitions) which are the attractors of regular and chaotic dynamics respectively, 
together with a nontrivial fixed point associated to critical (null Lyapunov exponent) dynamics. 

\section{Type-II intermittency: definition and basic statistical properties}

For definiteness we chose the case of type-II intermittency \cite{schuster} as it develops 
for nonlinear iterated maps in the vicinity of a Neimark-Sacker bifurcation. 
As a canonical example of a discrete system exhibiting this type of dynamics let us 
consider the iterated complex map:
\begin{equation}
z_{t+1}=\alpha z_t+\mu |z_t|^{2}z_t
\end{equation}
with $\alpha=(1+\epsilon)e^{\varphi i}$ and $\mu \in \mathbb{R}$. If we rewrite the variable $z$ in its polar form $z=xe^{\theta i}$, 
we can decompose the dynamics of the system in a rotation of its argument $\theta$ and a nonlinear dynamics in its modulus $x$. 
We focus on the dynamics of the modulus $x$, which is where intermittency appears:
\begin{equation}
x_{t+1}=(1+\epsilon)x_t+x_t^{3}=F(x_t)
\end{equation}
with a choice of $\mu=1$ for simplicity.\\

\begin{figure*}[tbp]
\centering
\includegraphics[width=0.7\columnwidth]{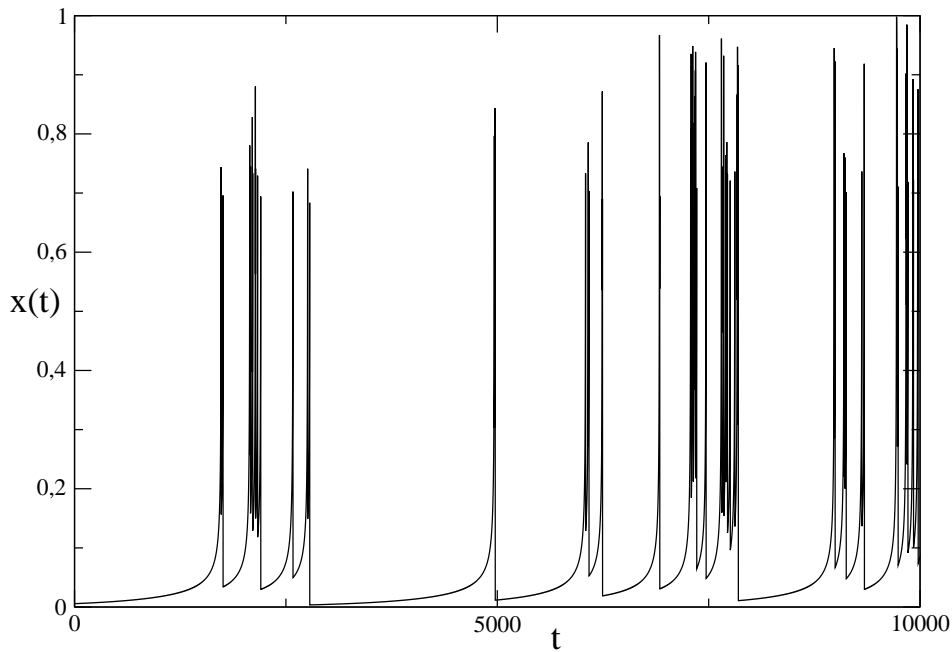}
\caption{A sample trajectory of the map defined in equation \ref{map}, with $\epsilon=10^{-3}$. 
In type-II intermittency, trajectories monotonously increase until reinjection takes place.  Reinjections close to the unstable fixed point take long journeys to depart from its neighborhood, and are experimentally seen as pseudoperiodic (laminar) phases. Chaotic bursts are actually concatenation of short laminar trends generated out from 
reinjections far from the unstable fixed point.}
\label{series}
\end{figure*}

\noindent This one dimensional iterated map has an unstable fixed point in $x=0$ for $1\ggg\epsilon>0$. 
If no additional constraints were imposed the map would diverge for initial conditions larger than zero, however, 
we can bound the phase space by introducing a modular congruence in the definition of this map  
\begin{equation}
x_{t+1}=(1+\epsilon)x_t+x_t^{3}\ \mod 1.
\label{map}
\end{equation}
The trajectories of this map are monotonically increasing functions up to a certain value $0<x_r<1$ that fulfills $1=(1+\epsilon)x_r+x_r^3$. 
Then, the modular congruency reinjects it \textit{somewhere} in the vicinity of the unstable fixed point $x^*=0$, where they remain for a 
\textit{certain} time (a certain number of iterations) $t$ until they escape, go beyond $x_r$ and are reinjected once again (see figure 
\ref{series} for a graphical illustration). Reinjections close to the unstable fixed point take long journeys to depart from its neighborhood, 
and are experimentally seen as pseudoperiodic (laminar) phases. Chaotic bursts are actually concatenation of short trends generated out from 
reinjections far from the unstable fixed point (see in figure \ref{series} the alternation between long laminar phases and chaotic bursts, 
governed by the location of the reinjection value).\\
Note that there is also a second value $x_r<x_{2r} < 1$ very close to $1$ that fulfills $2=(1+\epsilon )x_{2r}+x_{2r}^3$ from which the trajectories 
are also reinjected in the vicinity of $x=0^+$. As a numeric guide, for $\epsilon=10^{-3}$ 
$x_r=0.68204\dots$ and $x_{2r}=0.99975\dots$. This map densely fill the phase space [0,1] and evidence sensitivity to initial conditions 
for $\epsilon>0$, regular dynamics for $\epsilon<0$ \cite{schuster} and criticality at $\epsilon=0$. $\epsilon$ actually determines the 
distance of the system to the bifurcation.\\

\noindent A paradigmatic feature of type-II intermittency is the scaling of the mean length of the laminar trends 
$\langle\ell\rangle \sim \epsilon^{-1}$. This scaling is suggested by assuming the laminar 
trends to start at a $x_0\lll 1$ close enough to the fixed point of our map ($x^*=0$), such that
\begin{equation}
x_{1}=(1+\epsilon)x_0+x_0^{3}\approx(1+\epsilon)x_0,\ x_0\lll 1.
\end{equation}
This lead us by recurrence to 
\begin{equation}
x_{\ell}\approx(1+\epsilon)^\ell x_0=[1+\epsilon\ell+{\cal O}(\epsilon^2)]x_0\approx(1+\epsilon\ell) x_0 
\end{equation}
with $x_{\ell}$ the value at which we can consider the laminar trend is over. From the former expression we get:
\begin{equation}
\ell\approx\frac{1}{\epsilon}\left(\frac{x_{\ell}}{x_0}-1\right)
\label{l_x0}
\end{equation}
and averaging $\ell$ for different initial values $x_{0_i}$ of the trend, we get
\begin{equation}
\langle\ell\rangle \approx \lim_{n\rightarrow \infty}\frac{1}{n}\sum_{i=1}^{n}\frac{1}{\epsilon}\left(\frac{x_{\ell}}{x_{0_i}}-1\right)=
\left(\left\langle{x_0^{-1}}\right\rangle x_{\ell}-1\right)\epsilon^{-1},\ \left\langle{x_0^{-1}}\right\rangle=\lim_{n\rightarrow \infty}\frac{1}{n}\sum_{i=1}^{n}\frac{1}{x_{0_i}}.
\end{equation}\\ 

\begin{figure*}[tbp]
\centering
\includegraphics[width=0.7\columnwidth]{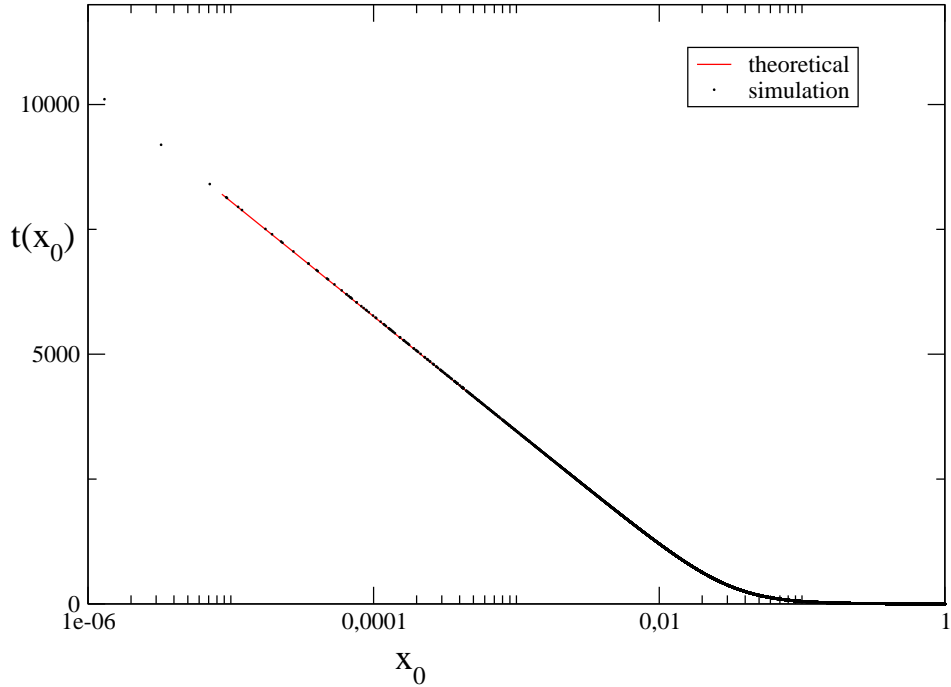}
\caption{Reinjection time $t(x_0)$ for the map defined in equation \ref{map}, with $\epsilon=10^{-3}$ (see fig. \ref{series}). 
Red curve stands for the theoretical expression deduced in eq. \ref{t_x0}. Black circles are direct numerical measures in a trajectory 
of length $T=10^7$.}
\label{t_x0_fig}
\end{figure*}

\noindent However, note that the definition of a laminar trend is somewhat ad-hoc, as we shall define what we consider to be such a phase, 
that is, when do we consider that the trajectory has escaped laminarity. Accordingly, the mean length of laminar trends 
$\langle\ell\rangle$ is not a variable that directly contains information particularly relevant from the point of view of 
the graph theoretical description as we shall later see. A more objective measure is the time $t$ between reinjections 
(or its mean value $\langle t\rangle$). An expression for this variable $t$ as a function of the initial value $x_0$ can 
be found by taking the continuous limit and traducing our map to a flow:
\begin{equation}
x_{t+1}-x_{t}\approx \frac{dx}{dt}=x(\epsilon+x^2)\Rightarrow \int_{x_0}^{x_r}\frac{dx}{x(\epsilon+x^2)}=\int_{t_0}^{t}dt.
\end{equation}
Assuming a relatively small initial time ($t-t_0\approx t$) this integration is straightforward and lead us to the following expression:
\begin{equation}
t(x_0,\epsilon)=\frac{1}{\epsilon}\log{\frac{x_r\sqrt{\epsilon+x_0^2}}{x_0\sqrt{\epsilon+x_r^2}}}
\label{t_x0}
\end{equation}
where $x_0$ is the initial or so called reinjection value (see fig. \ref{t_x0_fig}). If, on this expression, we make the substitution $x_r=x_{\ell}\ll 1$ 
we can recover by approximation the expression found in eq. \ref{l_x0}. A probability density function $g(t)$ can now be deduced 
for the reinjection time $t$ from the corresponding pdf $f(x_0)$ of the reinjection values $x_0$, since we know that:
\begin{equation}
g(t)dt=f(x_0)dx_0\Rightarrow g(t) = f[x_0(t)]\left|\frac{dx_0(t)}{dt}\right|.
\label{g_t_th}
\end{equation}
From eq. \ref{t_x0} we have that 
\begin{equation}
x_0(t,\epsilon)=\frac{\epsilon^{1/2}}{\sqrt{ae^{2\epsilon t}-1}},\ a=\frac{\epsilon+x_r^2}{x_r^2}. 
\end{equation}
In absence of a more detailed description, we may assume in a first approximation that $f(x_0)$ is reasonably uniformly 
distributed in the interval [0,1] which takes the role of the phase space of our system ($x_0\rightarrow U[0,1]\Rightarrow f(x_0)=1$). 
This yields the following density:
\begin{equation}
g(t)=f[x_0(t)]\left|\frac{dx_0(t)}{dt}\right|=a\epsilon^{3/2}\frac{e^{2\epsilon t}}{(ae^{2\epsilon t}-1)^{3/2}}\approx
\left\{
\begin{array}{ll}
\frac{a\sqrt{\epsilon}}{2}t^{-1},\ t_m \gg t \\
\frac{\epsilon^{3/2}}{\sqrt{a}}e^{-\epsilon t},\ t_m \ll t,
\end{array}
\right.
\label{g_t}
\end{equation}
where $t_m$ is a characteristic time scale for which we can consider that the function 
behaves as a power law for short times $t$ ($t_m\gg t$) and exponentially 
for long times $t$ ($t_m \ll t$).
If we calculate the mean value $\langle t\rangle$
with this approach for short and long times, we obtain:
\begin{equation}
\langle t\rangle=\int_{t_0}^{\infty}tg(t)dt\approx \frac{a\sqrt{\epsilon}}{2}\int_{t_0}^{t_m}dt+\int_{t_m}^{\infty}\frac{\epsilon^{3/2}}{\sqrt{a}}te^{-\epsilon t}dt=\nonumber
\end{equation}
\begin{equation}
=\frac{at_{m}}{2}\epsilon^{1/2}+\frac{t_me^{-\epsilon t_m}}{\sqrt{a}}\epsilon^{1/2}+\frac{e^{-\epsilon t_m}-1}{\sqrt{a}}\epsilon^{-1/2}\sim \epsilon^{-1/2},
\label{t_med}
\end{equation}
\begin{figure*}[tbp]
\centering
\includegraphics[width=0.7\columnwidth]{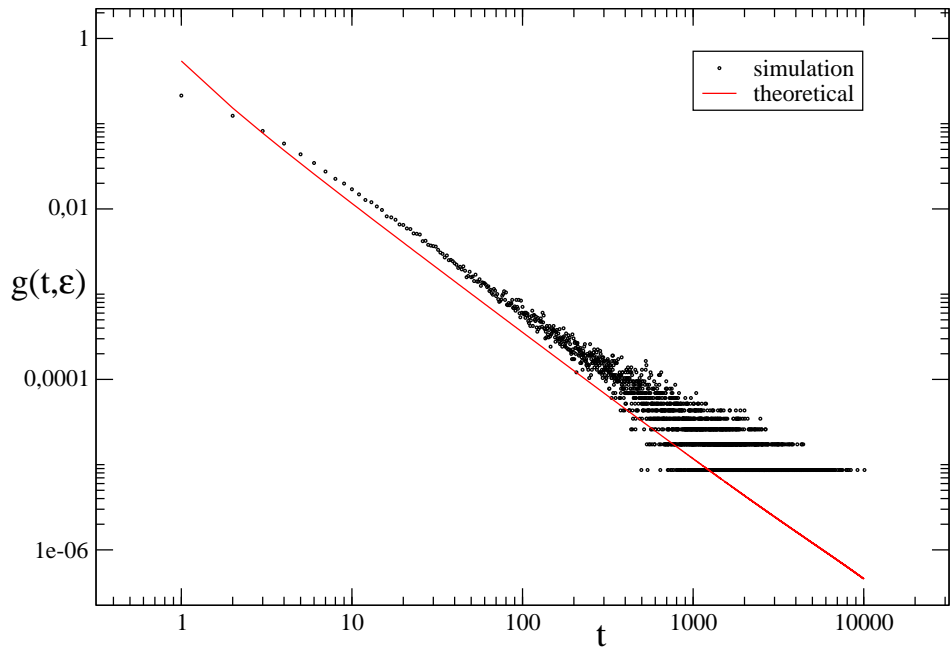}
\caption{Log-log plot of the probability density function of times to reinjection $g(t;\epsilon=10^{-3})$ for the map defined in equation \ref{map}. 
Red curve stands for the theoretical approximation. Black dots are extracted from the simulation of a trajectory of leght $L=10^{7}$.}
\label{g_t_fig}
\end{figure*}
which exhibits a scaling with $\epsilon$ that however quantitatively differs from the one found for $\ell$ (again, $t_{m}-t_{0}\approx t_{m}$). 
We can recover the scaling for $\ell$ with this approach just by assuming that, 
in order a trend to be considered as laminar, the reinjected value has to be 
in the vicinity of the fixed point: $x_0\in(0,b],\ b\sim\sqrt{\epsilon}$ or otherwise, 
$f(x_0)$ must be a reasonably uniform pdf: $x_0\rightarrow U[0,\sqrt{\epsilon}]\Rightarrow f(x_0)=1/\sqrt{\epsilon}$,   
The pdf, $g(\ell)$, that we obtain in this case is slightly different:
\begin{equation}
g(\ell)=f[x_0(\ell)]\left|\frac{dx_0(\ell)}{d\ell}\right|=2\epsilon\frac{e^{2\epsilon \ell}}{(2e^{2\epsilon \ell}-1)^{3/2}}\approx \left\{
\begin{array}{ll}
\ell^{-3/2},\  \ell_m \gg \ell \\
\frac{\epsilon}{\sqrt{2}}e^{-\epsilon \ell},\ \ell_m \ll \ell
\end{array}
\right.
\end{equation}
This new pdf predicts a scaling for the mean length of the laminar trend of the form: 
\begin{equation}
\langle \ell\rangle=\int_{\ell_0}^{\infty}\ell g(\ell)d\ell\approx \int_{\ell_0}^{\ell_m}\ell^{-1/2}d\ell+\int_{\ell_m}^{\infty}\frac{\epsilon}{\sqrt{2}}\ell e^{-\epsilon \ell}d\ell= \nonumber
\end{equation}
\begin{equation}
=2\sqrt{\ell_m}+\frac{\ell_me^{-\epsilon\ell_m}}{\sqrt{2}}\epsilon+\frac{e^{-\epsilon\ell_m}-1}{\sqrt{2}}\epsilon^{-1}\sim \epsilon^{-1}.
\label{l_med}
\end{equation}
where $\ell_m-\ell_0\approx \ell_m$. A comparison of these expressions with the results of numerical simulations is shown in figure \ref{g_t_fig}.
Note that, if we expand our interval of initial conditions in the reinjection to the whole phase space $[0,1]$, 
we recover $t$ ($\lim_{b\rightarrow 1}\ell=t$). 

\section{Transformation of intermittent time series into Horizontal Visibility graphs}

The Horizontal Visibility (HV) algorithm \cite{luqueetal2009} assigns
each datum $x_{t}$ of a time series $\{x_{t}\}_{t=1,2,...}$ to a node $n_t$ in its 
associated HV graph (HVg), where $n_t$ and $n_{t'}$ are two connected nodes if 
$x_{t},x_{t'}>x_{\tau}$ for all $\tau$ such that $t<\tau<t'$. The resulting are 
outerplanar graphs connected through a Hamiltonian path \cite{gutin} whose structural properties capture the statistics enclosed in 
the associated series \cite{review_graphs}. A relevant measure is the degree distribution $P(k)$, 
that accounts for the probability of a randomly chosen node to have degree $k$, 
which has been showed to encode key dynamical properties such as fractality, chaoticity 
or reversibility to cite some \cite{review_graphs}.\\

\begin{figure*}[tbp]
\centering
\includegraphics[width=0.9\columnwidth]{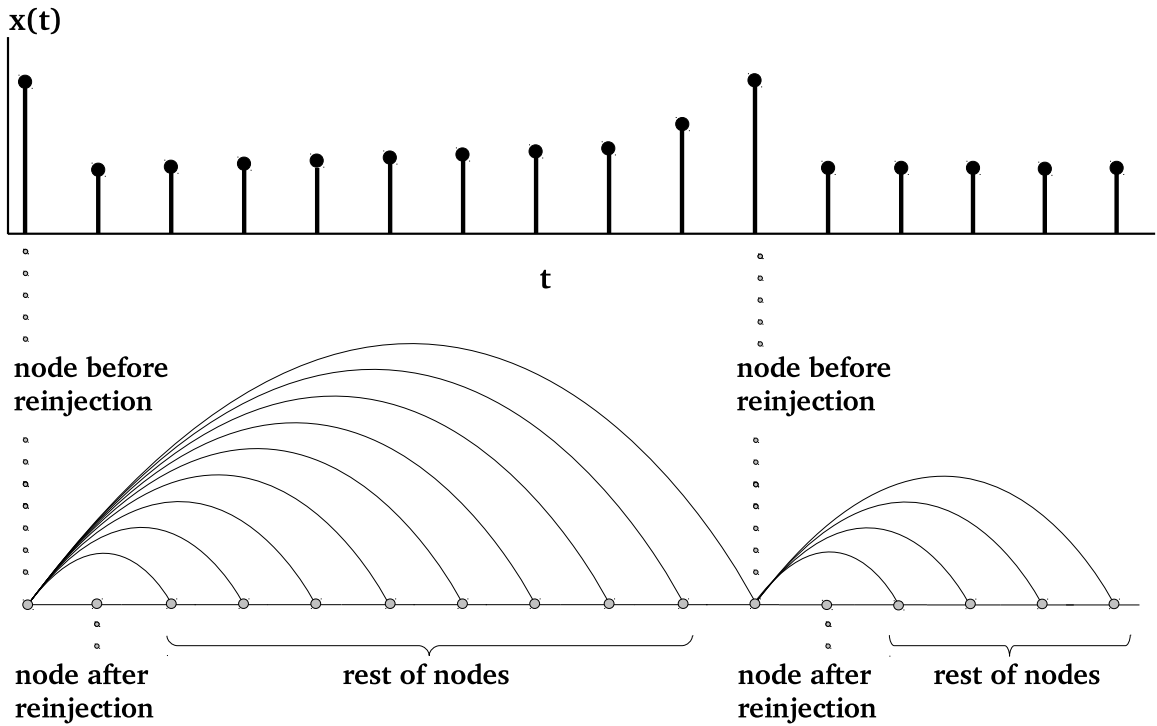}
\caption{Sketch of an intermittent trajectory along with its associated HV graph. 
Nodes 'before reinjection' correspond to values in the trajectory which 
have 'visibility' over the values until the next reinjection. Nodes just after 
reinjection correspond to values in the trajectory which are bounded by two 
higher values.}
\label{cartoon}
\end{figure*}

For illustrative purposes, in figure \ref{cartoon} we represent a sketch of a type-II intermittent 
series along with its associated HV graph. At odds with the phenomenology of type-I intermittency, 
whose mapping consisted of several 
repetitions of a $T$-node motif (periodic backbone of period $T$ associated to the ghost of the period 3 series) linked to the first node of the 
following laminar trend and interwoven with the chaotically connected nodes associated to the 
chaotic bursts between laminar trends \cite{interm_I}, the case of type-II lacks any periodic backbone or chaotic burst, and  
consists of nodes associated to reinjections and nodes linked to the last node of the 
previous reinjection. That is, the method naturally distinguishes reinjections from each other 
and do not include the (nonetheless ambiguous) distinction between laminarity and burstiness. 

Accordingly, we can classify nodes in three 
different categories (see fig. \ref{cartoon}):\\ 
a) nodes located just before a reinjection $n_{r{-}}$, whose degree distribution will be later discussed in detail,\\ 
b) nodes located just after a reinjection $n_{r{+}}$, with a trivial degree distribution ($P_{r^{+}}(k=2)=1$) and\\ 
c) the rest of the nodes $n_t$, whose degree distribution is also trivial ($P_{t}(k=3)=1$).\\

\noindent Based on these observations, in what follows we derive some topological properties of these graphs 
and will show that they indeed incorporate the main statistical properties of type-II intermittency. 

\subsection{Degree distribution $P(k)$}
Consider the degree 
distribution $P(k)$, that describes the probability that a randomly chosen node of a graph has $k$ links (degree $k$). 
The previous features allow us to decompose the degree distribution of type-II intermittency graphs  
as a weighted sum of the aforementioned contributions:
\begin{equation}
P(k)= f_{r}P_{r^{-}}(k)+f_{r}P_{r^{+}}(k)+(1-2f_{r})P_{t}(k),
\label{p_k_eq}
\end{equation}
where $f_r$ is the reinjection fraction  
\begin{equation}
f_r=\lim_{\tau\rightarrow\infty}\frac{n_r}{\tau},
\end{equation}
and $n_r$ is the number of reinjections that have occurred up to time $\tau$.

\begin{figure*}[tbp]
\centering
\includegraphics[width=0.7\columnwidth]{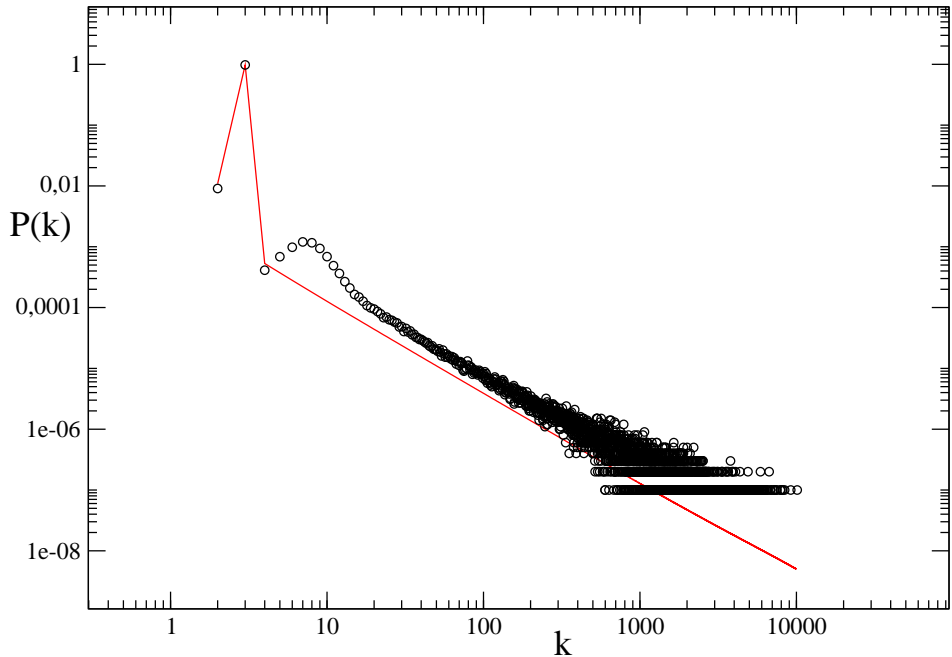}
\caption{(Circles) Log-log plot of the degree distribution $P(k)$ of the Horizontal Visibility graph (HVg) 
mapped from a trajectory generated by equation \ref{map} for $\epsilon=10^{-3}$. (Red curve) theoretical 
expression for the same distribution as presented in eq. \ref{p_k_eq}.}
\label{p_k_fig}
\end{figure*}

\noindent On the other hand, the degree of the nodes before a reinjection can be decomposed in two 
different contributions $k=t+k'$. The first term $t$ consists on the visibility the node has over 
the following laminar trend up to the next reinjection and it is distributed like 
$P_{r^-}^t(t)\approx g(t)$. 
The second term $k'=k-t$ comes from the visibility the node has over the rest of the reinjections 
and its distribution can be supposed to be the characteristic exponential distribution found in 
HVg's that come from stochastic/chaotic processes like a reinjection: 
$P_{r^-}^r(k)=\delta(k-k')e^{-\lambda k},\ k'\in\mathbb{Z}^+-\{ 1\}$ \cite{toral10}.
We have one further consideration: by construction, the degree of these nodes has 
an additional restriction: $k_{r{-}}\geq 4$. As an approach, we can assume $k\gg k'$ 
or, equivalently, $k\approx t$. 
Therefore, the degree distribution for these nodes turns out to be: 
\begin{equation}
P_{r^-}(k)\approx\int_{0}^{k}P_{r^-}^t(t)P_{r^-}^r(k-t)dt=g(t),\ k=t
\end{equation}
In figure \ref{p_k_fig} we plot in log-log the approximate theoretical expression for 
the degree distribution $P(k,\epsilon)$, for a concrete value of $\epsilon=10^{-3}$, 
along with the numerics.

\subsection{Variance $\protect\sigma _{k}^{2}=\langle k^{2}\rangle
-\langle k\rangle ^{2}$}

\begin{figure*}[tbp]
\centering
\includegraphics[width=0.7\columnwidth]{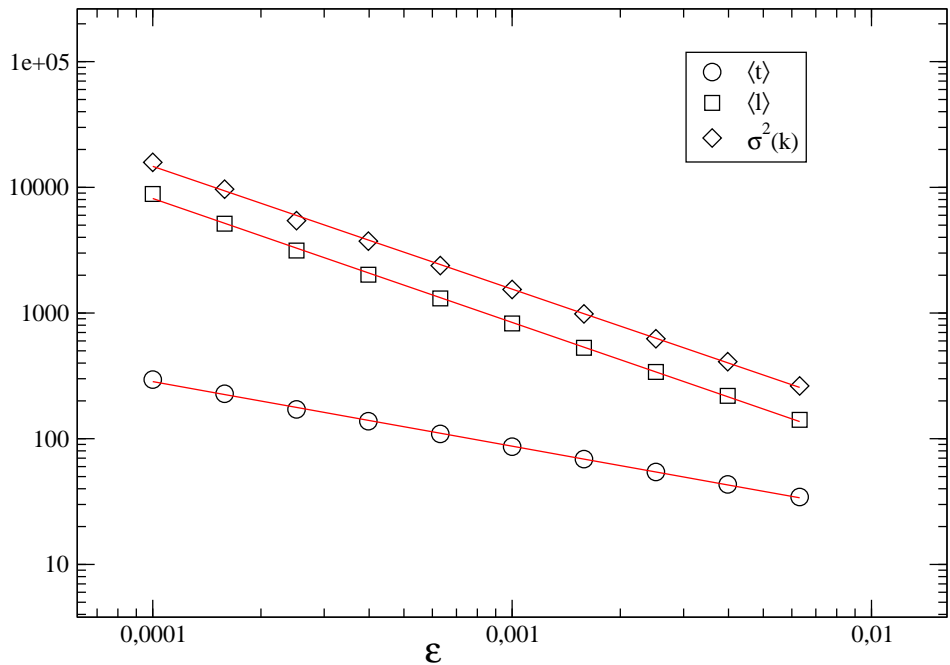}
\caption{Log-log plot of $\langle t\rangle$ (circles) and $\langle \ell\rangle$ (squares) as a function of $\epsilon$, 
numerically calculated from time series of $10^7$ data. Diamonds correspond to the variance of the degree distribution 
of the associated HV graph (numerical results). In each case, solid lines are the predicted analytical scaling 
$\langle t\rangle\sim\epsilon^{-1/2}$ (eq. \ref{t_med}), $\langle \ell\rangle\sim\epsilon^{-1}$ (eq. \ref{l_med}) and 
$\langle k^2\rangle-\langle k\rangle^2\sim\epsilon^{-1}$ (eq. \ref{varrrr}), in good agreement with the numerics.}
\label{t_l_var_k}
\end{figure*}

In this section we are going to verify our previous hypothesis $P_{r^{-}}(k)\approx g(t)$ 
by deducing an expression for the mean value of the degree distribution $\langle k\rangle$, 
and we are going to analyze how the second order moment of the distribution  $\sigma_{k}^{2}$ inherits the scaling of 
$\langle \ell \rangle$. 
Let us start with the mean degree $\langle k\rangle$. Recall that HV graphs associated to generic aperiodic series tend, 
as size increases, to a constant mean degree $\langle k\rangle=4$ \cite{caos} and that HVgs are, by construction, 
connected graphs where node $i$ has degree 
$k\geq 2\  \forall i$ (they have a Hamiltonian path). Then,
\begin{equation}
\langle k\rangle=\sum_{k=2}^{\infty}kP(k)= (1-2f_{r})\cdot3+f_{r}\cdot2+f_{r}\sum_{k=4}^{\infty}kP_{r^-}(k)=4\Rightarrow \sum_{k=4}^{\infty}kP_{r^-}(k)= 4+f_{r}^{-1}=\langle k_{r^-}\rangle. 
\end{equation}
This expression tell us that the mean value of the connectivity of the nodes before reinjetion 
$\langle k_{r^-}\rangle$ behaves as the inverse of the reinjection fraction $f_{r}^{-1}$. 
Moreover, we can express the reinjection fraction as:
\begin{equation}
f_r=1-\lim_{\tau\rightarrow\infty}\frac{n_r\cdot\langle t\rangle}{\tau}=1-f_r\cdot\langle t\rangle \Rightarrow f_{r}^{-1}=\langle t\rangle +1 \Rightarrow \langle k_{r^-}\rangle=\langle t\rangle +5
\label{fr}
\end{equation}
for which we have assumed that the number of nodes between reinjections is quite the same as 
the number of reinjections multiplied by the mean value of the reinjection time 
($N_t=n_r\cdot\langle t\rangle$), which gives a compact expression  
$\langle t\rangle=\langle k_{r^-}\rangle-5\sim \epsilon^{-1/2}$ that indicates the behaviour of  
$P_{r^-}(k)$ and $g(t)$ coincides up to first order, and that $f_r$ goes to zero as $\epsilon^{1/2}$.\\

\noindent If we now move to the second order moment, we get:    
\begin{equation}
\sigma _{k}^{2}=\langle k^2\rangle-\langle k\rangle^2=\sum_{k=2}^{\infty}k^{2}P(k)-16=f_{r}\sum_{k=4}^{\infty}k^{2}P_{r^-}(k)+7(1-2f_{r}), 
\label{sigma_k}
\end{equation}
and following our hypothesis we can approximate 
\begin{equation}
\sum_{k=4}^{\infty}k^{2}P_{r^-}(k)\approx\int_{t_0}^{\infty}t^{2}g(t)dt\approx \frac{a\sqrt{\epsilon}}{2}\int_{t_0}^{t_m}tdt +\int_{t_m}^{\infty}\frac{\epsilon^{3/2}}{\sqrt{a}}t^{2}e^{-\epsilon t}dt=\nonumber
\end{equation}
\begin{equation}
=\frac{at_m^2}{2}\epsilon^{1/2}+\frac{2}{\sqrt{a}}\epsilon^{-3/2}\sim \epsilon^{-3/2},\nonumber
\end{equation}
where the last formula requires small values of $\epsilon$ and $t_m-t_0\approx t_m$. As $f_r\sim \epsilon^{1/2}$ 
for small values of $\epsilon$, in this low limit we have a leading order of 
\begin{equation}
\sigma _{k}^{2}\sim \epsilon^{-1}
\label{varrrr}
\end{equation}
which reproduces the expected scaling  (see figure \ref{t_l_var_k} for a comparison with numerics). We conclude 
that the scaling of the mean length of laminar phases with $\epsilon$ is inherited in network space by a similar 
scaling in the variance of the degree distribution.

\subsection{Scaling of Lyapunov exponent: Block entropies $h_n$}

\begin{figure*}[tbp]
\centering
\includegraphics[width=0.7\columnwidth]{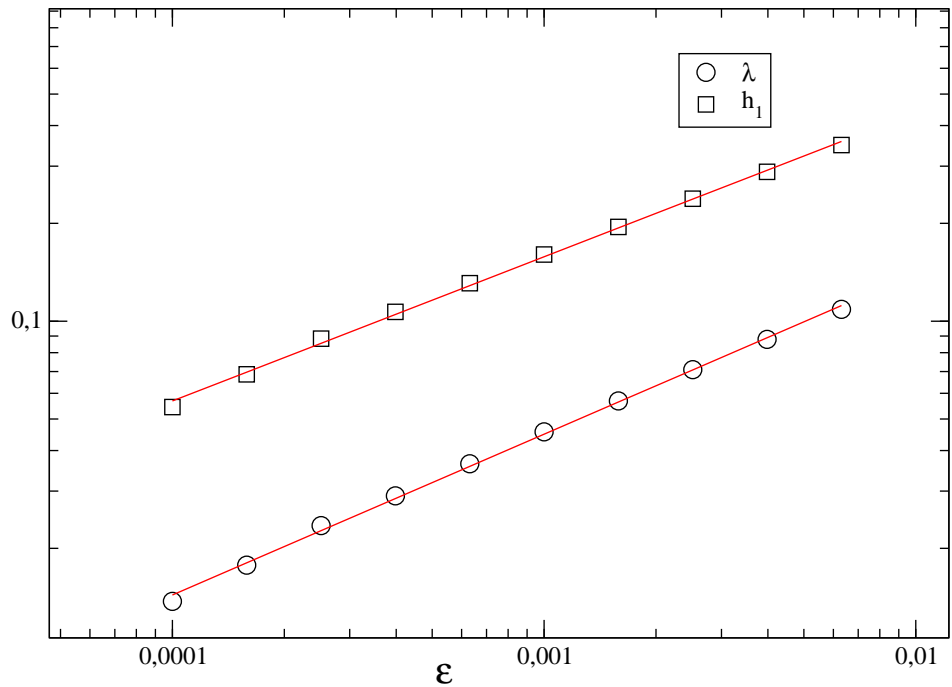}
\caption{(Circles) Log-log plot of the Lyapunov exponent $\lambda$
calculated numerically from a trajectory of length $L=10^{7}$ with the expression 
$\lambda=\lim_{t\rightarrow\infty}\frac{1}{t}\sum_{i=0}^{t-1}\ln|F'(x_{i})|$. Solid line is a regression with the 
predicted scaling: $\lambda(\epsilon)\sim \epsilon^{1/2}$. (Squares) Log-log plot of the block-$1$ graph entropy 
$h_{1}(\epsilon)$ (see the text) of the Horizontal Visibility graph associated to the same trajectory along with 
a regression, whose best power law fit reads $h_1\sim \epsilon^{0.44}$.}
\label{lyap_h1}
\end{figure*}

The second paradigmatic feature of type-II intermittency is the scaling of the Lyapunov exponent $\lambda$ 
(which for a map $x(t+1)=F(t)$ reads $\lambda=\lim_{t\rightarrow\infty}\frac{1}{t}\sum_{i=0}^{t-1}\ln|F'(x_{i})|$) 
with respect to the distance to criticality $\lambda(\epsilon)\sim \epsilon^{1/2}$ \cite{schuster}. 
Note that Lyapunov exponents characterize a purely dynamical feature and, 
although some graph-theoretical extensions of these exponents have been recently advanced 
\cite{pla}, it is not evident at all how to cast this dynamical behaviour into a 
graph-theoretical realm. However, note that
Pesin identity relates positive Lyapunov exponents of 
chaotic trajectories with Kolmogorov-Sinai rate entropy in dynamical systems. Based on this identity, a relation between 
Lyapunov exponents of maps and Shannon-like entropies over the degree distribution of the associated visibility graphs has 
been proposed \cite{plos,caos,pla,interm_I}. The block-$1$ graph theoretical entropy $h_1$ \cite{interm_I} is defined as 
$$h_1=-\sum_{k=2}^{\infty}P(k)\log P(k),$$ and block-$n$ entropies take into account of higher order statistics of blocks 
$P(k_1,k_2,...,k_n)$.\\

An approximate leading order calculation for $h_1$ can be performed assuming $P_{r^-}(k)\approx g(t)$ and recalling that 
the rest of the terms in $P(k)$ only contribute for $k=2,3$. $h_1$ reduces to a linear combination
\begin{eqnarray}
-h_1&=&f_{r}\log{f_{r}}+(1-2f_{r})\log{(1-2f_{r})}+\sum_{k=4}^{\infty}f_{r}P_{r^-}(k)\log{[f_{r}P_{r^-}(k)]} \nonumber\\
&\approx& f_{r}\log{f_{r}}+(1-2f_{r})\log{(1-2f_{r})}+\int_{t_0}^{\infty}f_{r}g(t)\log{[f_{r}g(t)]}dt \nonumber\\
&=&2f_{r}\log{f_{r}}+(1-2f_{r})\log{(1-2f_{r})}+f_r\int_{t_0}^{\infty}g(t)\log{[g(t)]}dt 
\end{eqnarray}
In order to derive the leading order of $h_1$ in $\epsilon$, we recall that $f_r\sim\epsilon^{1/2}$ for small values of $\epsilon$.
Let us pay attention to the last integral, which can be evaluated by a time scale separation for short ($t_m \gg t$) 
and long times ($t_m \ll t$) as in (\ref{g_t}): 

\begin{equation}
\int_{t_0}^{\infty}g(t)\log{g(t)}dt\approx \int_{t_0}^{t_{m}}\frac{a\sqrt{\epsilon}}{2}t^{-1}\log{\frac{a\sqrt{\epsilon}}{2}t^{-1}}dt+\int_{t_m}^{\infty}\frac{\epsilon^{3/2}}{\sqrt{a}}e^{-\epsilon t}\log{\frac{\epsilon^{3/2}}{\sqrt{a}}e^{-\epsilon t}}dt
\end{equation}
\begin{eqnarray}
&=&\frac{a\sqrt{\epsilon}}{2}\log{ \frac{ a\sqrt{\epsilon}}{2}}\int_{1}^{t_{m}}\frac{1}{t}dt-\frac{a\sqrt{\epsilon}}{2}\int_{1}^{t_{m}}\frac{\log{t}}{t}dt
+\frac{ \epsilon^{3/2} }{\sqrt{a}}\log{ \frac{\epsilon^{3/2}}{\sqrt{a}}}\int_{t_{m}}^{\infty}e^{-\epsilon t}dt+\frac{\epsilon^{3/2}}{\sqrt{a}}\int_{t_{m}}^{\infty} t(-\epsilon e^{-\epsilon t})dt\nonumber\\
&=&\frac{a\epsilon^{1/2} }{2}\log{ \frac{a\epsilon^{1/2}}{2}}\log{t_m}-\frac{a\epsilon^{1/2}}{2}(\log{t_m})^{2}
+\frac{\epsilon^{1/2}}{\sqrt{a}}\log{ \frac{\epsilon^{3/2}}{\sqrt{a}}}e^{-\epsilon t_m}+\frac{\epsilon^{1/2}}{\sqrt{a}}e^{-\epsilon t_m}+\frac{t_{m}\epsilon^{3/2}}{\sqrt{a}}e^{-\epsilon t_m}.
\end{eqnarray}

After a brief inspection, we conclude that for small values of $\epsilon$, the leading term of $h_1$ is not exactly 
$\sim\epsilon^{1/2}$ but $\sim-\epsilon^{1/2}\log{\epsilon}$, what should yield an approximate scaling albeit with 
a smaller exponent. In figure \ref{lyap_h1} we plot in log-log the numerical results of $h_1$ as a function of $\epsilon$, 
together with the numerics of $\lambda(\epsilon)$ in the same range ($\epsilon \ll 1$). As expected, $h_1$ shows a 
reasonably similar scaling, although with a slightly smaller exponent (the numerical fit is $\alpha\approx 0.44$).\\ 

\begin{figure*}[tbp]
\centering
\includegraphics[width=0.7\columnwidth]{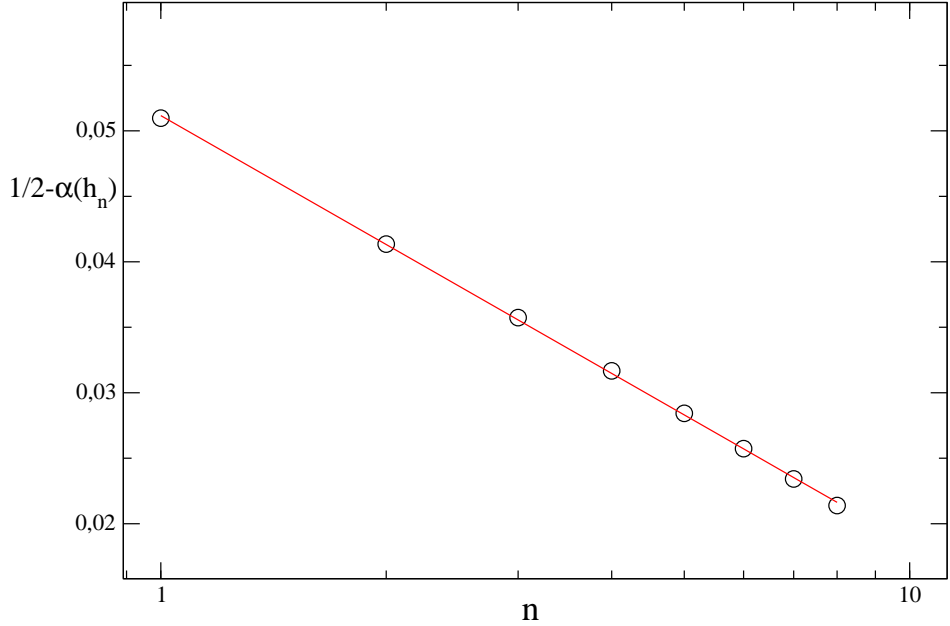}
\caption{(Circles) Semilog plot of the function $(1/2-\alpha(n))$ as a function of $n$, where $1/2$ 
and $\alpha(n)$ are the respective scaling exponents of the Lyapunov exponent and the block entropy 
$h_n$ with respect to $\epsilon$. We observe that these differences decreases logarithmically with $n$, 
suggesting the presence of a graph theoretical Pesin identity (see the text).}
\label{lyap-h_n}
\end{figure*}

\noindent If we now proceed as in 
\cite{interm_I} and we increase the size $n$ of the block in the graph-theoretical entropy $h_n$:
\begin{equation}
h_{n}=-\frac{1}{n}\sum_{k_{1},...,k_{n}}P(k_{1},...,k_{n})\log P(k_{1},...,k_{n}).
\end{equation}
we numerically observe a tendency in the scaling behaviour $h_n(\epsilon)\sim \epsilon^{\alpha(n)}$ 
that suggests $\lim_{n\rightarrow \infty}\alpha(n)\longrightarrow 1/2$ and therefore points towards 
a graph theoretical Pesin identity $$\lim_{n\rightarrow \infty} h_n=\lambda.$$
Numerical evidence for these latter results are shown in figure \ref{lyap-h_n}.

\section{Graph-theoretical renormalization group analysis}

As in previous works \cite{plos,caos,quasi,interm_I}, we can define a Renormalization Group (RG) transformation 
$\mathcal{R}$ on the HV graphs $G$ as the coarse-graining of every couple of adjacent nodes where at 
least one of them has degree $k=2$ into a block node that inherits the links of the previous two 
nodes. In other words, the operation removes form the graph every node of degree $k=2$ along with its two links. 
Assuming infinitely long series (in order to avoid the rescaling procedure in standard RG), in what follows we 
argue that the flow induced by iteratively performing this RG operation classifies dynamics coming from 
above and below transition, although the phase portrait turns to be very different from the one found in type-I 
intermittency \cite{interm_I}.

\begin{figure*}[tbp]
\centering
\includegraphics[width=0.7\columnwidth]{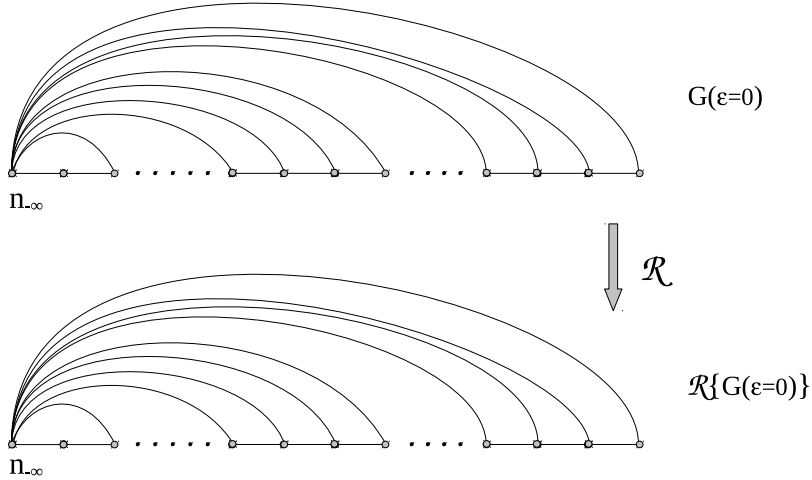}
\caption{Cartoon of an HV graph at criticality $G(\epsilon=0)$ (mapped from a trajectory at the onset of chaos in a Neimark-Sacker bifurcation, 
$\epsilon=0$ in equation \ref{map}). By construction the graph is invariant under renormalization (see the text), such that 
$\mathcal{R}\{G(\epsilon=0)\}=G(\epsilon=0)$. As any perturbation in $\epsilon$ generates a RG flow that takes the HV graph away from this 
equilibrium towards the stable attractors (parsimoniously towards the random graph $G_{\text{rand}}$ for positive perturbations and 
instantaneously to the chain graph $G_0$ for negative perturbations), $G(\epsilon=0)$ constitutes an unstable fixed point of the 
Renormalization Group flow associated to critical dynamics on the map and acts as a boundary between regular ($\epsilon<0$) and 
chaotic ($\epsilon>0$) dynamics.}
\label{RG}
\end{figure*}

\begin{itemize}
\item When $\epsilon<0$, trajectories are damped and rapidly converge to a constant series $x(t)=0 \ \forall t$, 
so after a transient the associated HVG $G_{0}$ will be a chain graph with $k=2$ for all nodes \cite{plos}.
A graph which is indeed the RG attractor of regular dynamics, invariant under renormalization $\mathcal{R}\{G_{0}\}=G_{0}$ 
\citep{caos} and represents a minimum for the entropy functional $h_1=0$.

\item When $\epsilon>0$ (the situation mostly addressed in the former sections), the associated HVGs describe 
a flow in RG phase space where, in each iteration of the renormalization procedure, the so called 
'nodes after reinjection' ($n_{r^{+}},\ k=2$) are eliminated progressively and, step by step, the laminar phases disappear, 
leaving essentially a series of 'nodes before reinjection' which form mainly an uncorrelated random series. 
The analogous flow in graph space slowly converges towards $G_{\text{rand}}$, the universal HV graph associated 
to an uncorrelated series, which constitutes a stable fixed point of the RG flow  
$\lim_{p\rightarrow \infty} {\cal R}^{(p)}\{G(\epsilon>0)\}=G_{\text{rand}}={\cal R}(G_{\text{rand}})$ \cite{caos}. 
As it has been proved that $G_{\text{rand}}$ is a maximally entropic HV graph \cite{caos}, we conclude that the RG 
flow $\forall\ \epsilon>0$ increases the entropy of the system as it breaks correlations.

\item At the bifurcation ($\epsilon=0$) the trajectories are monotonically increasing series 
bounded at $-\infty$ by a large value (the ghost of the intermittent regime), which maps into 
a HV graph which is itself indeed invariant under renormalization $G_{c}=G(\epsilon=0)=\mathcal{R}\{G(\epsilon=0)\}$ 
(see figure \ref{RG} for a graphical illustration). Its degree distribution can be described for a graph of $N$ nodes 
(where the limit $N\rightarrow\infty$ will be taken afterwards) as $P(k=2)=P(k=N-1)=1/N; P(k=3)=(N-2)/N$, yielding 
also a null entropy $h_1=0$ in the limit of diverging sizes. This constitutes the third fixed point of the
RG flow. Note that this fixed point is different from $G_0$, nonetheless, they both share the same entropy. 
The reason is that there is no actual parsimonious trajectory connecting $G(\epsilon=0)$ to $G(\epsilon<0)=G_0$, 
once you apply an infinitisemal perturbation making $\epsilon<0$, after a (large transient) you reach the 
stationary damped solution $x(t)=0\ \forall\ t$, whose associated HVg is directly $G_0$. There is hence no RG flow or  
entropy production associated with the path. On the other hand, any positive perturbation in $\epsilon$ is 
amplified in the RG flow, taking the graph to  $G_{\text{rand}}$. We conclude that $G(\epsilon=0)$ is an 
unstable fixed point of the RG flow. 
\end{itemize}

\section{Conclusions}
To conclude, in this work we have further advanced the Horizontal Visibility graph theory by providing an analytical 
and numerical graph theoretical description of type-II intermittency route to chaos, extending previous results for 
type-I intermittency \cite{interm_I}. We have shown that the key ingredients of type-II intermittency (concrete scalings 
of the mean length of laminar trends and Lyapunov exponent with respect to $\epsilon$) are recovered in the network 
realm by comparable scalings of the variance of the degree distribution and Shannon block entropy respectively. Finally, 
we have recasted the problem into a graph-theoretical renormalization group framework and have shown that the 
graph-theoretical fixed points of the RG flow distinguish regular, chaotic and critical dynamics. Whereas the trivial 
fixed points of the RG flow are universal attractors of regular and chaotic dynamics respectively, the unstable 
graph-theoretical fixed point, the RG flows and the associated entropy production paths are unique for type-II intermittency, 
being for instance a different scenario than what we found  in type-I intermittency \cite{interm_I} 
or in other routes to chaos \cite{caos, quasi}.\\

\noindent \textbf{Acknowledgements.} We acknowledge financial support from UPM project AL13-PID-09.


\end{document}